# REFLECTIONS, LEARNINGS AND PROPOSED INTERVENTIONS ON DATA VALIDATION AND DATA USE FOR ACTION IN HEALTH: A CASE OF MOZAMBIQUE

Nilza Collinson, University Eduardo Mondlane, University of Oslo, nilzac@ifi.uio.no

Zeferino Saugene, University Eduardo Mondlane, University of Oslo, zeferino.saugene@uem.ac.mz

Jørn Braa, University of Oslo, jornbraa@gmail.com

Sundeep Sahay, University of Oslo, sundeeps@ifi.uio.no

Emilio Mosse, University Eduardo Mondlane, emosse@uem.mz

**Abstract:** The ideal of a country's health information system (HIS) is to develop processes that ensure easy collection of relevant data and enable their conversion to useful health indicators, which guide decision making and support health interventions. In many Low- and Middle-Income Countries (LMICs), actively engaged in health reform efforts, the role of HIS is crucial, particularly in terms of quality of data and its ability to inspire trust in decision makers to actively use routine HIS data. Recognizing digital platforms potential to support those efforts, several interventions have been implemented in many LMICs. In turn, while the transition from paper registers to digital platforms carries the promise of improving data quality processes, this promise has been notoriously complex to materialize in practice. The authors draw upon more than 15 years of experience implementing HIS in Mozambique to understand how the potential of digital platforms have been realized with respect to data quality, what are the gaps and required remedial steps.

## 1. INTRODUCTION

One of the drivers for huge investments to Health Information Systems (HIS) implementation is the recognition of power of data to improve health care related informational practices. According to Lohr (2015, p.60) "data should be the vital raw material that strengthens and improves the machinery of decision making". However, "the data paradox is that a world richer and richer in data has so far yielded little payoff in most fields" (Lohr, 2015, p.60), including in health. Thus, while the supply of data races ahead, its use lags badly for various social-technical and political reasons. In this paper, we explore how this process has unfolded over time in Mozambique, a LMIC country that has, since the 90's, been using digital platforms to report and analyze health data (Braa et al, 2001). We explore developments and changes in the national HIS, resulting from digital interventions around the implementation of a digital platform known as the District Health Information System (DHIS), aimed at strengthening health care informational practices.

The efforts to introduce DHIS in Mozambique initiated early in 2000s, with its first desktop-based version, after its successful development and implementation in South Africa. The focus of this introduction has been on decentralization, while also attempting to gradually empower local authorities at lower levels of the hierarchy to have more control over their actions (Chilundo, 2004). Although, several studies conducted during this initial implementation revealed multiple challenges to guaranteeing quality of data, including technical, social and organizational conditions (Braa et al, 2001; Chilundo, 2004; Mosse & Byrne, 2005). Despite initial efforts and subsequent interventions in the following years, DHIS2 only came to be successfully implemented in the country several years after and scaled up nationwide in 2016. However, the challenge of data use to strengthen interventions remain despite the increasing availability of digital data.

This study was conducted as a formal assessment exercise carried out in 2019 by a joint team of researchers from University of Oslo (UiO) from Norway and Eduardo Mondlane University (UEM) from Mozambique, including representatives from the Ministry of Health (MOH) and several actors from the health sector on the sites visited. We focused on data management processes around the national HIS platform, designated SISMA, which is developed on top of DHIS2, pursuing a deeper understanding on how it has contributed to strengthening data management processes, in particular on data quality.

The aim of this paper was thus to identify the nature of the existing challenges and discuss what remedial steps could be taken to improve the current situation. We report from the findings of this assessment exercise attempting to provide both conceptual and practical contributions that may help strengthen practices and better realize the promise of digital technologies for improving HIS.

## 2. LITERATURE REVIEW

A national HIS should support the production of data that can be converted to health indicators and, thus support decision making and improve health interventions. A key challenge to realize this ideal is to balance the definition of relevant data set elements, so as to optimize the workload of frontline health workers while supporting their care provision processes (Shaw, 2005). Additionally, information needs of other stakeholders such as programme managers, policy makers and donors also need to be adequately considered (AbouZahr & Boerma, 2005; Cibulskis, 2005).

Latifov and Sahay (2012, p. 1) argue for building a more synergistic relation between data and indicators, suggesting that the stronger the linkage "the stronger will be the actionability of the HIS". Thus, availability and quality of health data is crucial in improving this link between data and its actionability. In turn, use of data is intrinsically related with its quality, and the more it is used, will push the supply of quality data obtained (Braa et al, 2012).

Aside from their innumerous potential, Digital platforms come with the promise of improving availability and use of data. This potential has been exploited by several organizations, operating in different quadrants, including commercial and public sectors. We take as an example, the engagement of a long-term research and development programme called HISP (Health Information System Programme) in developing a free and open-source digital platform called the DHIS2, now in use in more than 100 LMIC contexts (Adu-Gyamfi et al, 2019). This digital platform was built with basic aims to support essential data collection and validation by lower levels in the HIS, and support calculation of actionable indicators, while also supporting other information systems (Shaw, 2005). The expression "information for local action" was born alongside this digital platform and has grown in "meaning and purpose" over time (Braa & Sahay, 2012, p. iv).

Despite potential of digital platforms several challenges pose as threats to their successful adoption, implementation and furthermore to thrive in same proportion in different settings. Roland et al (2017) discusses challenges faced to ensure their flexibility through design, and several approaches adopted to enable adequate involvement of actors intervening in the process. Participatory Design (PD) as, in this sense, been seen as key to ensure that digital platforms incorporate and adjust adequately to several contexts simultaneously and provide needed and expected support. Although, due to changeable dynamics in the settings where long term developments take place, different PD approaches may be needed. Accordingly, HISP's PD approaches has gradually changed over time, shifting from single PD to community PD (Roland et al., 2017). And, DHIS2 has improved in its design to incorporate "flexibility, extensibility, and modularity" (Adu-Gyamfi et al, 2019:5) to cope with ongoing changes between multiple development and implementations, and accommodate incoming requirements.

In turn, as suggested by Krishna & Walsham (2005) the success of strengthening intervention is intrinsically dependent on other than technical aspects of the tool itself, including several aspects coming from the social context. Practices embedded within the organization and other social systems from the setting, existing interests of several groups, inadequate education related to information technology (Krishna & Walsham, 2005), governance approaches for decision making and use of resources (Bevir, 2011), and many others, play a significant role. And, strategies incorporating "tailored and country-specific guidance" and capacity strengthening interventions (Adu-Gyamnfy et al, 2019: 5) may be needed. Even though, as Latifov & Sahay (2012) advises, it is not guaranteed that institutional changes occur immediately, as long-term socio-political negotiations within the HIS are involved. This makes it important for Information System (IS) research to go beyond the technical factors and expand the focus on understanding other social related reasons justifying and explaining why this is the case and how these impediments to data use can be better addressed.

## 3. METHODS

An action research approach, as advocated by HISP (Braa & Nielsen, 2015), have been adopted along the implementation of HIS in Mozambique: to understand the routines and practices of the health staff at every level of the system, and to identify the gaps and challenges related to data use. This study ran as an assessment exercise included in that continuous process. A mixed team was assembled consisting of researchers from UiO and UEM, representatives from the MOH and staff from the provinces studied. And, a combination of data collection techniques was used including field visits, interviews, observations and focus groups discussions. Four of the researchers involved as authors in this paper, participated actively in the pilot of DHIS in Mozambique and in subsequent processes until present date, although with different levels of interventions, inputting their personal and valuable experiences from the process into the study. These and the fifth researcher, all participated in the fieldwork visits to the sites in 2019, sometimes split in groups to reach different settings.

### 3.1. Data collection

The fieldwork, run in a period of one month, covering six provinces, including the three sites where the original pilot was conducted in the early 2000s, as described in Table 1.

| Provinces visited | criteria for site selection | Nr. Districts visited | Nr. Health Units Visited | Instruments & Processes Observed |
|---|---|---|---|---|
| Maputo City | Headquarters' site | 1 | | Technical support, Data management, training processes |
| Maputo Province | Closest to the Headquarters | 2 | 2 | Books, charts, reports and official documents & Data collection processes |
| Gaza | Pilot site | 3 | 1 | Books, charts, reports & a validation Meeting (validation process), a Statistical Meeting (presentation and use processes) |
| Inhambane | Pilot site | 3 | 2 | Books, charts, reports & a routine program Meeting (to discuss data entry, validation, entry, presentation and use processes) |
| Niassa | Pilot site | 4 | 2 | Books, charts, reports & Data collection and Data entry processes |
| Zambezia | Far from the Headquarters | 1 | 1 | Books, charts, reports |

**Table 1. Overview of the fieldwork: sites visited and processes observed**

During the site visits, the team had opportunity to participate in three routine meetings, including a validation routine meeting, a statistical meeting and a routine meeting from a health programme team. In all these meetings, the team observed and at the end engaged in discussions regarding issues reported, either related to technical aspects of the digital platform or to the use of other technological resources available for daily operations.

In all of these opportunities, the team was able to gather sensibilities from different stakeholders involved in data management processes. For instance, in the statistical meeting, which is organized at the District level and led by the head doctor of the District, the participants reached a number of 20 people including the head statistician from the district, representatives from several health programs and the head doctor of the main Hospital of the District.

During the visits to health facilities, the team discussed with doctors and technicians, while also observing the environment and practices in course was able to engage with the nurses during their activities and data collection. These visits were also used as an opportunity to consult some of the paper-based instruments (books and charts) being used to collect and present daily registries, and see some of the digital instruments received as part of support provided by particular programs to collect specific data. The overall fieldwork enabled the team to gather an overview of data related processes: collection, validation, entry (capture), presentation and use, while also to access various data reports and official documents that were included in the study. Table 2 provides a resume of this process.

| Data related Process (*What*) | Site (Where) | Data collection method (*How*) | Source of information (*From what*) |
|---|---|---|---|
| Collection | Health Facilities<br>District Health Management sector | Observation of process on site<br>Observation of instruments<br>Discussion with nurses, statistical and health program officials on site | Consult paper-based instruments (books, charts fixed on walls)<br>Picture capture<br>View of tablets used to collect data |
| Validation | Health Facilities<br>District Health Management sector | Observation of process in meetings<br>Discussion with statistical and health program officials on site | Participation in meeting<br>Annotations<br>Picture capture |
| Entry (Capture) | District Health Management sector | Observation of process on site<br>Online Questionnaire (spread by WhatsApp and email)<br>Observation of instruments<br>Discussion with administrator, Head Doctors, statistical and health program officials | Annotations<br>Picture capture<br>View of computers being used |
| Presentation and Use | District Health Management sector | Observation of process in meetings<br>Observation of instruments<br>Discussion with administrator, Head Doctors, statistical and health program officials | Participation in meeting,<br>Consult paper-based instruments (Reports)<br>Annotations<br>Picture capture |

**Table 2. Summary of the data collection process: methods and sources.**

### 3.2. Process of Analysis

Our data analysis was based on a social-technical perspective, which examined the interconnected nature of technical and social challenges influencing data use processes. The information gathered was systematically organized into categories of issues to be analyzed afterwards, and, by adopting a social-technical perspective, we searched for different aspects in need of revision either from the instituted procedures perspective or from the software platform design itself. Thus, taking into account that this perspective "recognizes data, its utilization and software designed to support data processing is a product of institutional processes and practices" (Latifov & Sahay, 2012, p.2).

The analysis process involved five steps described as follows. In step 1 we organized the data collected on the field by categories (themes) of data related processes linked directly to DHIS: entry (capture), validation and use. We also created two categories related to activities and stakeholders involved; capacity strengthening; and governance. Step 2 was basically to identify existing limitations and suggested interventions for each process and the third step was dedicated to interpret the results and summarize them into four key findings. The following step included the preparation of the field report, to be shared afterwards with different parts interested in the process. The fifth and last step was dedicated to present the report in DHIS2 annual meeting, in order to consolidate and enrich some of the approaches raised by the team.

### 3.3. Findings

Normally, at the District level the health sector includes in its structure a District Service of Health, Women, Social Action (SDSMAS). This unit is responsible to coordinate all the processes related to the HIS from the setting, and includes within its team a statistical officer and a representative officer for each health programme. In this section we provide the overall findings summarized around data entry, data validation, data use, capacity strengthening, and governance themes. Where, under each theme, we discuss some of the practices, existing limitations and suggested interventions. Where possible, examples are provided to give the basis for our interpretations.

**Data entry**

We were able to witness different experiences and practices during the data entry process in the Provinces visited. While in the majority of the Districts visited the responsibility for data entry is from the statistical officers, in some this responsibility is shared with the programme officers (Cuamba district, in Niassa province) and in others is also delegated to nurses (Bilene district, in Gaza province, and Cumbane district, in Inhambane province).

In all places visited, for these processes, the health workers rely on some kind of technological device such as computers, tablets and smartphones to capture and entry data into SISMA. Although, not all of these belonged to the SDSMAS, some were provided by programme partners and in most cases the officers used personal devices to ease and accelerate their work. It was also noticed that in all the provinces, personal smartphones were being used to interact with other colleagues, share data and information, particularly using WhatsApp functionalities. For instance, we learn that almost everywhere two or three work groups were created in different levels, provincial, district and technical. One example of WhatsApp use, to receive data and shorten distances, we take from Cuamba, where the epidemiologic data was being shared using this tool to speed up the reception of data, coming from Health facilities that were located far from the SDSMAS, to be entered in SISMA.

We also noticed that data was entered in DHIS2 only after the monthly validation had taken place, which was conducted all based-on paper. The implication of this was, one, the validation process became extremely time consuming and work intensive; two, this process meant none of the DHIS2 features for data validation (like validation rules and data ranges) were being used.

**Data validation**

Overall, we found the practice of data validation to be quite robust and taking place on a systematic routine basis. MOH has given guidelines on the timeframe for the preparation of the resumes from the books, validation of data, data entry and use of data. There were monthly meetings in every district for validating the data. This meeting took place in the district, and sometimes at the health facility. We witnessed such meeting at Bilene district taking place in Mazivila health facility.

Officials from the districts met with staff from district health facilities who had gathered there, representing different health programmes. The entire meeting was based on the paper forms which the staff carried, questions were raised, the staff attempted to answer them and, in some cases, a need was identified to make changes. Sometimes, if anomalies were identified in the data, a phone call would be made to the health facility to seek clarifications and make changes, as seen in the Caronga district (Niassa province) office. In another district from the same province (in Sanga), one of data entry officers who had a motorcycle, would often drive down to the more distant facilities, to collect the copies of the form that were to be submitted, to accelerate the reception of data to be discussed with other facilities on data issues, and make required correction. This practice was probably due to the large distances between the district office and the health facility.

The staff in many of the health facilities and districts told us that the time given for data entry in the DHIS2 was very small, and they should be given more time. Even though some other officials counter pose that this shortage of time was verified because the data received from the health facilities was first entered in excel files, validated in the meetings and later on introduced in DHIS2. No data was entered in DHIS2 prior to validation and then the time to the introduction becomes short. In their opinion, if the period is extended the officials would relax and the problem would remain. We were interested to understand why this process was used to validate data, and how we could strengthen this practice. In one district the users said that they believed that once the data was entered into the DHIS2, it would get locked, and also become visible to the levels above. To avoid this, they preferred to do all the validation process on paper, prior to data entry, and only once that was completed, they would enter the data into DHIS2. To ensure completion, we were told in Sanga that they have the standard list from MOH of the indicators they needed to report to on a monthly basis. The district responsible would then compare the data reported from the facilities with the MOH list, to identify gaps and take corrective action. In some cases, Excel sheets were used for the data validation process. They use the Excel and target approach both to review data before capturing in DHIS2 and to discuss and analyze the data. The procedure includes: collect and summarize monthly data by facility and then enter in Excel, then look up data from the same month last year in DHIS2 and enter that manually. We saw how in Chibuto district (Gaza province) staff used tables with colored cells, red when under and green on or above the target, and discussed the achievements and possible causes with the facilities representatives, followed by exchange of personal experiences and lessons learned on the ground.

We found nearly all provinces to have at least two sets of institutionalized processes of meetings. One, the monthly meeting for data validation, described previously, and the second was a data use related meeting also taking place monthly, called Statistical Meeting, where the head Doctor of the District would meet with the district facilities heads and district managers of the health programs, to show the monthly data, present its analysis, identify areas of improvement, and provide recommendations for action. When we went to Mazivila health facility, we encountered several officials gathered in a Validation Meeting. On site, without many commodities, different officers representing health programs were gathered to discuss data reports received from all the district facilities. In turn, the Statistical Meeting occurring at the SDSMAS in Chibuto and Cuamba districts was organized in a more formal manner. We participated in the meeting in Chibuto and observed pictures and official documents from the last, as they ran in the same period.

We were also told that apart from these two monthly meetings, there was also a third meeting taking place which is for administrators. However, we did not get an opportunity to witness such a meeting. In addition to the monthly meetings, there were also quarterly meetings being held to look at the data over the time period at the provincial level and annual meetings led by the central level to share experiences and good practices between the provinces and define new strategies for the sector.

From what we perceived, though these different processes unfolded in different ways, the underlying theme evident was that health data is being taken seriously, and is seen as a basis for identifying action.

**Data use**

During the visits, we learnt that the conduct of these monthly meetings has been mandated by MOH for all provinces, and having seen these meetings taking place in all the sites visited, we infer it is relatively well-established practice nationally. We also found all districts and provinces visited to have incorporated relatively systematic processes around data use, even if they were manual or based on Excel. Although, nearly no districts or provinces were using DHIS2 visualization tools.

We were invited to participate in the Statistical Meeting in Chibuto district as guests. The meeting started with the presentation of Dashboards produced in PowerPoint displaying graphics with arrows and balloons with legends, highlighting comparisons of data from previous year and current year for the same period. For each health program, data were presented in tables with monthly resumes, with some data extracted from DHIS2 and tables formatted in Excel sheets. Facility results were highlighted denoting increases and reductions, emphasizing the results for each health program. After this initial presentation, the head Doctor discussed the results: questioning, demanding for answers and justifications, while also providing recommendations for action and improvement.

The question arises why this is the case and what can we do about it. Many reasons were identified, such as DHIS2 visualizations being seen as rigid and inflexible, inability to combine text and charts in Word document reports and PowerPoint presentations, infrastructure limitations and inadequate capacity.

**Capacity strengthening**

The Health sector has a mobility policy that allows and promotes reallocation of the professionals in this sector within the different provinces. While at the same time faces the challenges of losing professionals that shift to the partners, go for studies or even leave the sector. Those shifts are probably the biggest challenges for the statistical offices in the province, where the staff is constantly shifting and new staff comes to fill the gaps left behind.

We saw in almost every District that we visited new staff in the statistical sector that did not have participated in any specific training in using DHIS2, being only capacitated while at work performing the daily tasks. For example, in Maxixe district (Inhambane province), the three officials were newcomers and in Metarica district (Niassa province), the only official of the District learning at work. In this sense, even for the district head Doctors the training would be very relevant, as they stand as the lead of statistical departments at district and provincial levels. When we talked to the Head Doctor in Metarica this was one of his requests and according to him, one of his "crazy dreams" to have a device with DHIS2 that would allow him to monitor the indicators from the district on his hands.

**Governance**

We had the opportunity to see in the field the existing attempts to have qualified people to deal with data analysis and prepare information in a comprehensive manner to the health workers in the settings, while easing this responsibility from the other health workers. Although, these efforts are seen as being insufficient due to the limitations to hire public personnel within the Health Sector, where priorities are to incorporate more people with curative training and skills. For instance, we take the example of Niassa province where few statisticians are allocated: at the health facilities level 4 statisticians are allocated to the four Rural Hospitals from the Province, at the District level only 3 of the 17 Districts have statisticians allocated to the District Nucleus of Statistics (NED), and the Provincial Directorate (DPC) at provincial level with only 1 Statistician. In the other NED, to deal with statistics the province recruits preventive medicine technicians, as they receive some training in statistics during their formative training.

In terms of technical support for the system, in most of the sites visited we were told that the technical support teams are centralized at the province level providing remote support and, eventually if conditions are settled, they go to the districts to perform corrective and technical specific tasks related to the equipment allocated there. Regarding the support to digital platform, it was mostly given at distance by the statistician located in the NEDs at District and Province levels, and if this support is shown to be insufficient, then the technical teams allocated at the Ministry are involved. Regarding other system related issues, such as registry of new users and configuration of services and health facilities, interventions are conducted centrally at the MOH. Further support is also provided locally by an external team, supported and linked to the digital platform owner under the agreement and strict coordination with the Ministry.

## 4. DISCUSSION

While the current design of the HIS in Mozambique still unfolds in existing four levels of management, it is being perceived to have improved: from the ways data is being collected, to the flows of information till the availability of data to action. Though, the current assessment enabled the research team to observe and identify both positive outcomes and some aspects that may need further interventions. The key findings from this research study, so far, are summarized as follows:

1) DHIS2 primarily used for data storage and not for data use: The rationale behind the successful adoption of the DHIS2 adoption by LMICs has been its determination on providing data to empower lower levels. This has received very little attention by many health information systems. The current assessment has demonstrated how DHIS2 has democratized access to data in Mozambique. Despite these potentials we have observed that the system is primarily used as a data storage. The vast analytic tools developed to facilitate data visualization are barely used by lower levels. Instead DHIS2 users prefer to download data to excel and make charts and tables. These could have been easily generated from within DHIS2 system. Main problems are related to visualization features that are not flexible enough or not fitting use cases. MoH and district level people are using Excel and not DHIS2 for analyzing DHIS2 data. This makes it recognize the importance of functionalities that makes offline use of data possible, having data 'locally' is important and the key reason why people prefer Excel. Therefore, offline apps such as 'My Datamart', where data can be analyzed while offline, is much needed.

2) Continued use of existing tools of Excel and paper: Since DHIS2 is not being used to prepare data reports discussed either in the routine meetings or in the statistical meetings, we discovered also that it is not being used for data output. Instead, other tools are playing this role and providing these functionalities. The design of such generic instruments needs to be adjusted to local needs, and sometimes providing adjusted templates for reports that are automatically updated or introducing better ways to produce and share useful district and facility dashboards would make dashboards more useful. We acknowledge in this process, the need to also understand how the indicators are being produced, as for instance in Mozambique the standard is to use the targets as denominators, which is not supported in DHIS2 and poises as a must to be included.

3) Ambiguity in what population data to use: One aspect that should be recalled as well is related with the indicators being used as key to promote good data use and achieve intended goals. As such, we identified that in Mozambique at district level key indicators are being used to monitor data reported since the Facility level. In this case, adjustments in the instrument should be configured accordingly, introducing facility-based indicators with the inclusion of parameters such as, for instance, Facility population. We have seen that good district level data use relies upon indicators by facility, which again relies upon facility-based target population. Being able to compare facilities – monitoring facility-based indicators - is important for using data at district level. It is important to mention that the source for these parameters is already available in Mozambique, as facility populations are distributed and included at district level. These facility populations, however, are not included in DHIS2. In fact, population data is not included at all making it difficult to use DHIS2 for calculating coverage indicators.

4) Poor mechanisms of technical support, particularly in remote areas: This study revealed another issue that is constantly underestimated, provision of support and continuous capacity building for the practitioners in the field. In order to keep the system flowing there is a need to establish stronger bridges between different teams providing support at all levels of the HIS. The focus should be on local levels, in particular to districts and facilities, maintaining field visits on a routinely basis to better understand local needs and requirements, as part of the participatory design approach. For instance, the DHIS2 teams could easily have noticed that targets were used as indicator denominators by visiting a district and a province, but this was not done before. More generally; assessment of system status should be implemented as seen from the field not only from headquarters and needs to be carried out continuously, as part of a continuous participatory approach. Continuous capacity building should be encouraged within the teams supporting the implementations of any software platform. Space and time for recycling and update of knowledge is critical to maintenance of the systems. At the moment, for instance, we identified weaknesses in this component in the field, as the supporting team is facing difficulties managing metadata and keeping their database 'clean'. Ultimately to reduce some of these routine tasks and provide additional support to the teams to sensible tasks the system may incorporate this as functionalities. Indeed, these types of initiatives are in place and an app for help managing metadata is being developed by providing a way to manage and navigate between data sets, services and data elements. In DHIS2 categories of metadata are viewed as lists with a hierarchy within them, but not within them, making it difficult to manage and clean the metadata.

## 5. REFLECTIONS AND CONCLUSION REMARKS

In Mozambique, DHIS2 has been implemented and it is being used in a different manner from the initial intents, to support data evaluation, data visualization and data use. These probes the flexibility of the tool but reveals another insight to the developers that need to rethink the needs to take the whole workflow and social system in the targeted context into account. Here, some approaches such as the User participation and Participatory design are perceived to meet the task and need to be applied along the development and implementations. The information system is best understood as a social system – not a technical system. DHIS2 implementation has so far been carried out as a 'technical' implementation of software and needs to change to address the whole targeted context of use; the social system and relevant work practices.

Moreover, routine data use meetings in place are key to institutionalization of good data use. These meetings occur on a routine basis following a monthly calendar scheduled at the national level and followed at district and province level, using data collected at facilities and are seen as key to institutionalized data use practices. We perceive more generally, that data use indeed needs to become a natural part of all routine procedures such as routine data review, planning and M&E meetings and various quarterly and annual meetings and reporting procedures. We also found that DHIS2 was not customized to support these meetings and data use situations with appropriate dashboards, score cards and reporting templates. In this scenario we oversee a potential insight from a practice to the redesign of the instrument in such a way to support and improve these data use practices.

In general, any software platform is required to be flexible in various forms to accommodate adjustments and incorporate solutions to identified issues in the actual design. We have seen from the case that some of the main problems reported during the field visits are related to metadata management and difficulties in cleaning the database, making it problematic to run the system. In such cases, if the system does not provide internal functionalities, one option would be to suggest the incorporation of external apps developed to provide those functionalities inside DHIS2, like metadata management and metadata navigation apps for instance. This made possible also, through the existing portfolio of external apps being developed in other countries, which can be reviewed to elicit design ideas and requirements.